# Dynamic topological exciton-polaritons enabling ultrafast logic operations

Feng Jin[1], Hao Zheng[1], Zhe Zhang[1], Jiaohao Ren[1], Yuxi Liu[1], Qing Zhang[2], Daniele Sanvitto[3], Timothy C. H. Liew[1], Baile Zhang[1] and Rui Su[1,4,*]

[1]Division of Physics and Applied Physics, School of Physical and Mathematical Sciences, Nanyang Technological University, Singapore 637371, Singapore

[2]School of Materials Science and Engineering, Peking University, Beijing 100871, China

[3]CNR NANOTEC, Institute of Nanotechnology, via Monteroni, Lecce 73100, Italy

[4]School of Electrical and Electronic Engineering, Nanyang Technological University, Singapore 639798, Singapore

*E-mails: surui@ntu.edu.sg (R. S.)

**Abstract**

Topological active materials have emerged as powerful paradigm bridging the discovery of exotic topological phases of matter with the development of functional topological devices. The recent extension of these material systems into dynamic regime, where topological properties can be actively manipulated at ultrafast timescales, promises unprecedented control over topological states and their functionalities. However, translating the static topological lasing signals into high-performance logic functions remain highly challenging, which imposes a far more stringent set of materials attributes. Here, leveraging the strong nonlinearity and pronounced spectral isolation of perovskite exciton-polaritons embedded in a Dirac vortex microcavity, we experimentally demonstrate the dynamic topological Majorana-like state polariton condensation with its ultrafast logic operations at room temperature. By actively coordinating pump and control beams in both spectral and temporal domain, we dynamically steer the topological polariton condensation process and demonstrate AND and NOT logic operations, achieving record extinction ratio (~20 dB), extremely low control fluence (~0.2 nJ/cm$^2$) and sub-picosecond response time (~500 fs). Our results expand the frontier of dynamic topology and establish a novel pathway towards robust, ultrafast, and reconfigurable on-chip polaritonic logic circuits.

**Introduction**

Topological states of matter have attracted sustained interest across multiple branches of physics owing to their remarkable resilience against disorder and imperfections[1-4]. Beyond their static manifestations, topology physics is increasingly transitioning towards dynamic regime, where topological properties are no longer immutable attributes of static band structures but can be actively manipulated or steered in the temporal domain[5,6]. In particular, active materials that harness light-matter interactions to achieve optical gain and nonlinear responses represent a natural arena for this transition[7], giving rise to emerging advances such as Floquet band engineering[8-10], optical-controlled Haldane model[11-13], and topological phase transitions[14]. At the device scale, the emergence of topological active materials has also radically revolutionized active photonic devices, which provide a compact and robust route to generate highly coherent topological lasing within various engineered topological structures[15-20]. Translating these static synchronized topological lasing signals into the dynamic regime would constitute a paradigm shift, unlocking a novel generation of high-performance all-optical topological logic devices with enhanced robustness and operational stability arising from dynamical topology. Despite this transformative promise, such capabilities remain largely uncharted as of now, as they are fundamentally constrained by the limited light-matter coupling strength and nonlinear interactions in conventional active materials.

Exciton-polaritons, hybrid light-matter quasiparticles arising from the strong coupling between semiconductor excitons and cavity photons, offer a distinctive materials platform to resolve this dilemma[21-23]. As an unconventional active medium, exciton-polaritons unite the strong interparticle interactions of excitons with the agility of photons, giving rise to pronounced optical nonlinearities rooted in their inherently strong light-matter interactions[24-26]. With their driven-dissipative nature, they undergo nonlinear polariton condensation under bosonic stimulation[27-31], forming macroscopic quantum states with lasing emission that can be dynamically amplified or steered through optical seeding[32-36]. Particularly, recent advances in nanofabrication and emerging semiconductor emitters, such as inorganic halide perovskites and organic fluorescent proteins, have propelled static polariton condensation into diverse non-trivial topological contexts at room temperature[37-40]. These attributes position exciton-polaritons as a compelling platform for advancing dynamic topology and developing ultrafast, high-performance topological polaritonic logic architectures[23]. However, their experimental realizations remain highly challenging, primarily due to the narrow spectral isolation in most existing topological material platforms and the absence of reliable dynamic control over the topological polariton condensation.

In this work, we translate the topological polariton condensation into the dynamic regime and experimentally demonstrate a new generation of active topological logic gates within a perovskite Dirac vortex microcavity at room temperature. Distinct from previously reported topological contexts governed by band topology[15,39,41-43], our system represents a novel class of real-space topological defects, which hosts a topological Majorana-like state[20,44,45] pinned at mid-gap within a ~21 meV topological bandgap. Under a non-resonant pulsed excitation, we realize spontaneous exciton-polariton condensation into this Majorana-like state at room temperature. Harnessing the strong nonlinearity of perovskite exciton-polaritons together with the exceptionally large mid-gap spectral isolation, we dynamically amplify and suppress the topological polariton condensation at ultrafast timescale by coordinating pump and control beams in both spectral and temporal domains. Notably, this dynamic control mechanism enables the effective implementation of topological AND and NOT logic gates operating on the Majorana-like state, exhibiting sub-picosecond response times of ~500 fs and operating with extremely low control fluences down to 0.2 nJ/cm$^2$. Moreover, the topological AND gate and NOT gate achieve the extinction ratios of 20 dB and 16.4 dB, respectively, which establish the highest on-off contrasts reported to date for polaritonic logic.

**Results**

**Design of the topological perovskite Dirac vortex microcavity**

Figure 1a illustrates the schematic and operating principle of the dynamic topological exciton-polariton condensation with its logic functions. Specifically, under a non-resonant excitation, the driven-dissipative nature of exciton-polaritons allows them to achieve spontaneous condensation into the topological state with maximum net gain. By injecting an additional control beam from the back of the device as a seed, this condensation pathway can be steered in distinct ways. For instance, when the control beam is resonant with the topological condensation state, it could substantially reduce the condensation threshold and amplify the topological signal, thereby realizing a topological AND gate. Additionally, the control beam can also be tuned on resonance with the alternative trivial states to suppress and further switch off the spontaneous topological condensation, thus enabling NOT gates of the topological state. Despite this approach can, in principle, effectively steer the polariton condensation pathway, the experimental realization of polaritonic logic devices operating in the topological regime remains highly challenging. The main difficulty arises from the limited topological gap size and the resulting narrow spectral isolation in most existing topological structures. To overcome these limitations, we design and fabricate a perovskite Dirac vortex microcavity (Fig. 1a, see Methods), which hosts a topological Majorana-like state pinned at mid-gap within an exceptionally large topological bandgap. The core of this topological logic platform lies in the Dirac vortex lattice, serving as a polaritonic counterpart of the Jackiw-Rossi model[46]. In order to generate this lattice, we start from a pristine honeycomb lattice (lattice constant, $a = 1.10$ μm; micropillar radius, $R = 0.30$ μm), which supports Dirac-like dispersion with two valleys located at $\boldsymbol{K} = [\pm 4\pi/3a, 0]$. Next, we introduce a vortex-like Kekulé modulation into the honeycomb lattice geometry (Fig. 1b), described by the in-plane displacement

$$\delta \boldsymbol{D}(\boldsymbol{r}) = D(\boldsymbol{r})[\sin(\boldsymbol{K}\cdot\boldsymbol{r} + \theta(\boldsymbol{r})), \pm\cos(\boldsymbol{K}\cdot\boldsymbol{r} + \theta(\boldsymbol{r}))], \tag{1}$$

where $\boldsymbol{r} = (x,y)$ denotes the micropillar position in Cartesian coordinates, $\boldsymbol{K}$ is the valley wavevector and $\pm$ correspond to the A and B sublattices. The position-dependent displacement amplitude follows a radial profile of

$$D(\boldsymbol{r}) = D_0 \tanh(\boldsymbol{r}/\xi), \tag{2}$$

where $D_0 = 0.15a$ sets the modulation amplitude, and $\xi = 0.10a$ defines the radius of the undisturbed core. In addition, the position-dependent modulation phase $\theta(\boldsymbol{r})$ is defined as

$$\theta(\boldsymbol{r}) = w[\tan^{-1}(y/x) - \phi_0], \tag{3}$$

with the initial modulation phase offset $\phi_0 = \pi/6$, and the winding number encircling the vortex core $w = +1$, which makes the vortex a topological nontrivial defect. The inset of the Fig. 1c illustrates a hexagon supercell comprising sublattices A (orange) and B (purple), which are displaced from their original positions (indicated by black dashed lines) by a Kekulé modulation of amplitude $D = 0.15a$ and phase $\theta = \pi/3$. This modulation induces an intervalley coupling $\Delta \sim e^{i\theta(\boldsymbol{r})}D(\boldsymbol{r})$, folding the valleys to the $\Gamma$ point of the reconstructed Brillouin Zones (BZs) and opening a bandgap at the Dirac points. To probe the gap opening of the Kekulé modulation on the polaritonic honeycomb lattice, we present the band diagrams for all $2\pi$ values of modulation phase $\theta$ with the same amplitude (Fig. 1c) by solving the exciton-photon coupled Schrödinger equations (CSEs) in reciprocal space (see Methods). Inside the band diagrams, we observe a persistent vortex bandgap opening with a $2\pi/3$ angular periodicity owing to the $C_3$ symmetry of the lattice, which exhibits a maximum gap of ~32 meV at $\theta = \pi/3$ and a minimal gap of ~25 meV at $\theta = 0$ (Fig. 1c, Supplementary Note 1).

Different from the topological edge states governed by the topological bulk-edge correspondence, the Dirac vortex lattice can host a mid-gap topological Majorana-like state arising from the Jackiw-Rossi binding mechanism. Specifically, the hyperbolic tangent radial profile of Kekulé modulation $D(\boldsymbol{r})$ together with a non-zero winding number $w = +1$ produces a gapless core encircled by a gapped contour, which leads to the emergence of a topological mid-gap Majorana-like state confined in the middle of the $2\pi$ vortex gap at the lattice core. In order to validate this in our system, we construct a continuous model of the exciton-polariton Dirac vortex lattice with the CSEs (see Methods). The calculated polariton dispersion (left panel of Fig. 1d) and eigenstates (right panel of Fig. 1d) reveal a topological gap opening of ~21 meV (highlighted by black dashed lines), with the topological Majorana-like state pinned exactly at mid-gap and spectrally isolated from all other trivial states. Notably, this topological state is located at the $\Gamma$ points of the second BZs in momentum space (inset of Fig. 1d) and strongly confined at the vortex core to single type sublattices in the real space (Fig. 1e). Similar to the Majorana states in spinless p-wave superconductors[47], the polaritonic Majorana-like state is also protected by the particle-hole symmetry of the lattice, which makes this unique topological state robust and pinned at the mid-gap regardless of any defects as long as this symmetry is preserved (Supplementary Note 3). Benefiting from its large spectral isolation and robust protection, this Majorana-like state stands out as an ideal candidate for the realization of dynamic topological exciton-polariton condensation.

**Observation of the exciton-polariton topological Majorana-like state**

Building on this design, we fabricate the perovskite Dirac vortex lattice by patterning the ZEP520A spacer layer of the $CsPbBr_3$ perovskite planar microcavity. As shown in Fig. 2a, we present the atomic force microscopy (AFM) image of the fabricated perovskite Dirac vortex lattice, which clearly reveals the Kekulé modulation across the entire lattice profile. To examine the detailed morphology and uniformity of key regions, we characterize the AFM images of the magnified vortex core area (inset of Fig. 2a) and the bulk domains near the modulation phases of $\theta = \pi/3$ (left panel in Fig. 2b) and $\theta = 4\pi/3$ (right panel in Fig. 2b). All of these AFM images exhibit the precisely defined Dirac vortex lattice profile consistent with our design, together with excellent sample homogeneity, maintaining a thickness of ~ 60 nm at the lattice sites. To experimentally probe the emergence of the exciton-polariton topological Majorana-like state, we map the band structure via angle-resolved photoluminescence measurements. The entire perovskite Dirac vortex microcavity with a detuning of ~ -160 meV is non-resonantly excited by a 2.713 eV continuous-wave laser with a pumping spot diameter of ~ 35 μm at room temperature (see Methods). By collecting the emission from the bulk domains and the vortex core area, marked by the blue and orange translucent overlays in Fig. 2a, we observe distinct spectral features. Figure 2c and 2d illustrate the energy-wavevector dispersions collected from the bulk domain A (near $\theta = \pi/3$) and domain B (near $\theta = 4\pi/3$) along the $k_y$ direction, which reveals exceptionally large bandgaps opening of ~32 meV and ~25 meV, respectively.

Subsequently, we focus on the Dirac vortex core area and characterize its dispersion, which exhibits distinctly different band features, as shown in Fig. 2e. Owing to the 2π Kekulé modulation phase winding at the defect core, we observe the entire topological vortex bandgap opening of ~21 meV (outlined by the black dashed lines), in which an additional discrete mid-gap state emerges exactly at $E = \sim 2.257$ eV. This spectrally isolated state matches the predicted topological Majorana-like state (Fig. 1d) and holds an exceptionally large spectral isolation compared to previous polaritonic topological platforms (Supplementary Note 9). In addition, we validate the emergence of topological Majorana-like state by comparing the real-space photoluminescence profiles of the polaritonic Dirac vortex lattice at different energy levels with a 1 nm band-pass filter. As shown in Fig.2f, we observe that the emission from the Majorana-like state is strongly localized on single type sublattices in the lattice core, aligning well with the theoretical calculation (Fig. 1e).

Exciton-polariton condensation into the topological Majorana-like state underpins the realization of topological exciton-polariton logic gates, which serves as a robust logic operation state immune to imperfections. Thanks to their inherently driven-dissipative nature and limited lifetime, polaritons spontaneously condense into specific excited states with maximum net gain, which can be selectively determined by tuning the microcavity detuning. Here, to facilitate polariton condensation into the topological state, we employ a perovskite Dirac vortex microcavity sample with a detuning of ~ -130 meV, which is non-resonantly pumped by a pulsed laser at 3.1 eV with a pumping spot diameter of ~ 10 μm (see Methods). To experimentally probe the spontaneous condensation process into the topological Majorana-like state, we initially perform the momentum-space photoluminescence characterizations under varying pump fluence (0.7 $P_{th}$, 1.0 $P_{th}$, and 2.0 $P_{th}$). As shown in Fig. 2g, we present the dispersion of the perovskite Dirac vortex lattice under the pump fluence of 0.7 $P_{th}$ ($P_{th}$ = 12.2 μJ/cm$^2$), which exhibits similar band structure features as in the linear regime (Fig. 2c), but shifted to higher energy due to the more positive microcavity detuning. Increasing the pump fluence to $P = P_{th}$ (Fig. 2h), we observe an enhanced polariton occupation of the topological Majorana-like state, which marks the onset of the topological polariton condensation. When the pump fluence is further increased beyond the condensation threshold to $P$ = 2.0 $P_{th}$ (Fig. 2i), polaritons massively occupy the topological state, accompanied by pronounced linewidth narrowing and a significant intensity increase of nearly three orders of magnitude. The inset of Fig. 2i reveals a two-dimensional (2D) momentum-space image of polariton condensation into the topological Majorana-like state, where polaritons condense close to the *Γ* points of the second BZs, which is in good agreement with our theoretical calculations (the set of Fig.1d). In addition, the topological exciton-polariton spontaneous condensation process is further examined by real-space photoluminescence measurements (Supplementary Note 5). To quantitatively characterize the topological polariton spontaneous condensation process, we examine the evolution of emission intensity, linewidth, and peak energy extracted from the topological state as a function of pump fluence. As shown in the above panel of Fig. 2j, we observe that the integrated emission intensity of the topological state exhibits a pronounced nonlinear increase as the pump fluence exceeds the condensation threshold of $P_{th}$ = 12.2 μJ/cm$^2$, while the corresponding linewidth rapidly decreases from 4.27 to 2.10 meV. Concurrently, the peak energy of the topological state undergoes a continuous blueshift, resulting from the strong nonlinear repulsive interactions inherited from the excitonic component of exciton-polaritons (bottom panel of Fig. 2j). Collectively, these experimental observations validate the realization of spontaneous polariton condensation into the topological

Majorana-like state at room temperature.

**Realization of dynamic topological exciton-polariton condensation with its logic functions**

Exciton-polariton condensation inherently results from bosonic stimulation, where occupation of a state accelerates further scattering into it[24]. Introducing an additional control beam as a seed allows this stimulation pathway to be steered, thereby enabling reliable control over the condensation dynamics[26,33,34]. When resonant with the condensation state, the control beam could markedly reduce the condensation threshold and significantly amplify the emission signal, with the amplification most pronounced near the spontaneous condensation threshold, thereby enabling the realization of all-optical exciton-polariton AND gates. To experimentally realize a topological exciton-polariton AND gate, we employ a home-built optical setup (Fig. 3a) that allows precise control over both the spectral position of the control beam and its temporal delay relative to the pump beam (see Methods). With this configuration, we tune the pulsed control beam to the spectral position of the topological Majorana-like state in the same sample used above. As illustrated in Fig. 3b, the dispersion of the control beam nearly coincides with that of the topological state, occupying the momentum wavevector of $k_y = \sim 3.5$ μm$^{-1}$ and the energy level of $E = \sim 2.289$ eV, which is well separated from the trivial bulk bands. This alignment is further confirmed by comparing their integrated spectra fitted by Gaussian functions (Fig. 3c). To maximize the polariton condensation amplification, the control beam must be synchronized with the pump beam in the time domain (see Methods). Figure 3d presents the dispersions of the topological exciton-polariton condensation without and with the synchronized control beam of 10 nJ/cm² under a pump fluence of 15 μJ/cm², respectively. By comparing the dispersions, we observe a nearly order-of-magnitude amplification of the signal emission, which is further validated by the Gaussian-fitted integrated spectra in Fig. 3e.

Moreover, we compare the integrated intensity evolution of the topological state with and without control beam (constant at 10 nJ/cm²) under identical pump conditions (Fig. 3f). We observe that seeding the topological state with the control beam markedly reduces the condensation threshold from 12.2 to 8.6 μJ/cm², and amplifies the signal emission intensity across all pump fluences. The operating logic of our topological exciton-polariton AND gate is that the pump beam defines the topological Majorana-like state as the operation state, which can be gated by the control beam. We assign the emission intensity of the pump-only condition to logic level '0' and that with control beam seeding to logic level '1', respectively. The extinction ratio, given by the ratio between these two levels ('1'/'0'), serves as a central performance metric of the logic gates. As shown in Fig. 3g, we present the extinction ratio of our topological exciton-polariton AND gate as a function of pump fluence, which indicates that the extinction ratio peaks at ~ 20 dB as the pump approaches the spontaneous condensation threshold, establishing a record switching contrast for polaritonic logic gates (Supplementary Note 10). Under a fixed pump fluence of 12 μJ/cm², we further examine the dependence of extinction ratio on control beam fluence (Fig. 3h). We observe that the topological AND gate can be triggered with a control fluence down to 0.2 nJ/cm². Notably, this control fluence is nearly five orders of magnitude lower than the pump fluence, underscoring that our topological AND gate can operate with extremely low control energies while maintaining high extinction ratios. By modifying the time delay between the pump beam (12 μJ/cm²) and the control beam (10 nJ/cm²), we additionally probe the temporal response of our topological AND gate from '0' to '1' level. The integrated condensation intensities extracted from dispersions at different delays are normalized and plotted as a function of time delay (Fig. 3i). The resulting response dynamics, fitted with a Gaussian-convoluted exponential function, reveals a response time of ~ 490 fs at the given excitation conditions. Such ultrafast response dynamic behavior is further validated by the real-space photoluminescence measurements. At $\Delta\tau = 0$ ps, the real-space image exhibits the strongest polariton emission (top panel of Fig. 3i), whereas increasing the time delay to $\Delta\tau = 1.2$ ps results in a substantially reduced intensity, therefore it has been multiplied by a factor of 20 for better comparison (bottom panel of Fig. 3i).

In principle, spectrally aligning the control beam to be on resonance with alternative states enables the suppression and further switch off the spontaneous exciton-polariton condensation, effectively serving as a NOT gate[48]. Leveraging this mechanism, we demonstrate two NOT gates of the topological state within the same perovskite Dirac vortex microcavity sample. The schematic in Fig. 4a illustrates that the exciton-polaritons in the Dirac vortex microcavity spontaneously condense into the Majorana-like state in the absence of the control beam, which serves as the logic level '1' of the topological NOT gate. As shown in the left panel of Fig. 4b, we present the dispersion from the vortex core area of the same sample as used above in the linear region. Thanks to the Dirac vortex lattice design, the topological state is pinned to the middle of the topological vortex bandgap and spectrally well separated from the trivial band. This exceptionally large free spectral separation provides the basis for realizing topological NOT gates with high extinction ratios and negligible crosstalk. The right panel of Fig. 4b demonstrates the dispersion of the spontaneous topological exciton-polariton condensation at $E = \sim 2.292$ eV under a pump fluence of 18 μJ/cm², which reproduces the band features observed in Fig. 2i and Fig.3d. In addition, the real-space photoluminescence measurement further confirms that the polaritons spontaneously condense into the topological Majorana-like state (Fig. 4c). For better comparison, the intensities of both the dispersion and real-space image are multiplied by a factor of 2.

The fundamental operating principle of our topological NOT gate is to switch off the synchronized topological condensation signal (logic level '1') when the control beam arrives. Figure 4d and 4g schematically depict two distinct topological NOT gate configurations. When the control beam is resonant with either a higher- (Fig.4d) or lower- (Fig.4g) energy trivial state, as indicated by green arrows, the spontaneous topological exciton-polariton condensation is quenched and diverted into the trivial state condensation, corresponding to the logic level '0'. As shown in the right panels of Fig.4e and Fig.4h, we present the spectral positions of the control beam 1 and control beam 2, which are resonant with the higher- and lower-energy trivial states, respectively. Both control beams have the same fluence of 10 nJ/cm², and they are multiplied by a factor of 50 for visibility. Under a pump fluence of 18 μJ/cm², the corresponding dispersions of the polariton condensation driven by the control beam 1 and the control beam 2 reveal that the spontaneous topological condensation at $E = \sim 2.292$ eV (logic level '1') are switched off, while the higher trivial state at $E = \sim 2.307$ eV (Fig. 4e, right) and the lower trivial state at $E = \sim 2.281$ eV (Fig. 4h, right) are massively occupied, respectively. These transition behaviors from logic level '1' to '0' are further validated by the real-space photoluminescent images shown in Fig. 4f and 4i. Furthermore, we examine the transition from topological state (level '1') to higher-energy trivial state (level '0') as a function of control beam 1 fluence at a pump fluence of 18 μJ/cm². As shown in Fig. 4j, the emission from the topological state rapidly decreases once the control fluence exceeds 1.3 nJ/cm², indicating that the topological NOT gate switches from logic level '1' to '0'. When the control beam fluence is further increased beyond 10 nJ/cm², the suppression of the topological condensation stabilizes, and the extinction ratio reaches a maximum value of ~16.4 dB under these excitation conditions (Supplementary Note 7). As shown in Fig. 4k, we additionally probe the dynamic behaviors of this topological NOT gate from '1' to '0' level by varying the time delay between the pump beam (18 μJ/cm²) and control beam 1 (10 nJ/cm²), which reveals an ultrafast response time of ~510 fs. Our results collectively establish dynamic topological polariton condensation as a reliable mechanism for implementing both topological AND and NOT gates.

**Discussion**

In conclusion, by integrating the strong nonlinearity of perovskite exciton-polariton with the exceptional large spectral isolation of Dirac vortex microcavity, we have demonstrated the dynamic topological exciton-polariton condensation with its AND and NOT logic operations at room temperature. Notably, these topological logic devices outperform the previous polaritonic counterparts based on the planar microcavities, exhibiting record extinction ratio (~20 dB), ultrafast response time (~500 fs), and extremely low control fluence (down to 0.2 nJ/cm²). Our results not only establish a versatile approach to dynamically manipulate the topological exciton-polariton condensations towards highly robust topological logic devices, but also unlocks new avenues for exploring rich dynamical topology physics, such as Floquet Chern insulators, space-time topological events and topological pumping. While the logic operations demonstrated here correspond to Boolean logic, they are implemented in a Majorana-like state condensation that has been envisioned as a building block for topological quantum computing owing to its non-Abelian braiding nature[49,50]. This dual significance highlights the broader impact of our work, positioning perovskite Dirac vortex microcavity as platforms that not only enable robust polaritonic logic but also offer exciting opportunities for holonomic quantum information processing.

**Methods**

**Sample fabrication**

In our experiments, a perovskite Dirac vortex microcavity is used to realize the dynamic topological exciton-polariton condensation with its ultrafast logic operations. It consists of a bottom distributed Bragg reflectors (DBRs), polymethyl methacrylate (PMMA) spacer, an all-inorganic perovskite ($CsPbBr_3$) nanoplatelet, a ZEP520A spacer layer with Dirac vortex lattice, and top DBRs. Specifically, 12.5 pairs of $Ta_2O_5/SiO_2$ layers are deposited onto a glass substrate coated with 150 nm indium tin oxide (ITO) electrodes, fabricated by an electron beam evaporator (Cello 50D). A PMMA spacer layer with the thickness of 15 nm is then spin-coated onto the bottom DBRs, followed by the plasma surface treatment. The single-crystalline perovskite ($CsPbBr_3$) nanoplatelet with the thickness of 60 nm is grown via the chemical vapor deposition (CVD) method on the mica substrate. In detail, 0.2g $CsPbBr_3$ powder is placed in the center of the CVD chamber and the inside pressure is maintained at 42.5 torr with the $N_2$ flow rate of 30 sccm. The chamber is heated to 590°C within 5 min, followed by maintained for 10 min and cooled down to room temperature naturally. The perovskite layer is then transferred onto the bottom DBR substrate using a dry-transfer methods with Scotch tape. Next, a ZEP520A spacer layer with the thickness of 60 nm is spin-coated on the sample as the positive electron beam resist. It is patterned into the Dirac vortex lattice via standard electron beam lithography with an accelerating voltage of 30 kV, followed by Pentyl acetate development for 30 s and IPA rinsing for 30 s. Lastly, 10.5 pairs of $Ta_2O_5/SiO_2$ layers are deposited on it as top DBRs by the electron beam evaporator to complete the whole fabrication process.

**Optical spectroscopy characterizations**

The energy-resolved momentum-space and real-space spectrum are performed by a home-built angle-resolved spectroscopy setup with Fourier optics at room temperature. The polariton signal from the perovskite Dirac vortex microcavity is collected by a 50× objective lens (NA = 0.75) and sent to a 550 mm focal length spectrometer (Horiba iHR550) with a grating (600 lines/mm) and a liquid nitrogen cooled CCD (256 × 1024 pixels). For the photoluminescence measurements in Fig. 2, two different types of lasers are used to pump our sample, respectively. In the linear region, the polariton lattice is non-resonantly pumped by a continuous wave laser (2.713 eV) with a pumping spot diameter of ~35 μm. The real-space photoluminescence images at selected energies are measured by using a band-pass filter (linewidth, 1 nm; Semrock) in the detection path. In the nonlinear region, the sample is non-resonantly pumped by a pulsed laser (energy: 3.1 eV, pulse duration: 100 fs, and repetition rate: 1 kHz) with a homogeneous pump spot diameter of ~10 μm.

For the demonstration of the dynamic topological exciton-polariton condensation with its logic operations in Figs. 3 and 4, we employ a home-built optical setup that permits precise control over both the spectral position of the control beam and its temporal delay relative to the pump beam, as illustrated in Fig. 3a. Specifically, the femtosecond pulsed laser (energy: 3.1 eV, pulse duration: 100 fs, and repetition rate: 1 kHz) is split into two paths. In one path, the pulsed laser passes through a β-BaB$_2$O$_4$ (BBO) crystal to generate the pulsed pump beam at 3.1 eV via second-harmonic generation (SHG), which is then normally incident on the front side of the Dirac vortex microcavity sample. In the other path, the pulsed laser is sent to optical parametric amplification (OPA) system equipped with a band-pass filter (linewidth, 1 nm; Semrock) to produce a pulsed control beam at the target energy. Subsequently, after expansion by a 4$f$ optical system, the control beam is spatially filtered by a pinhole and then focused by a 50× objective lens (NA = 0.75) to be injected at an oblique angle from the back of the sample, which provides precise tunability of its location in the momentum space. In addition, the relative temporal delay between the control and pump beams can be finely controlled by a high precision moving stage.

**Theoretical calculations**

For modeling this system, we employ the coupled Schrödinger equations (CSEs) under the mean-field approximation. The CSEs used in this work are formulated as:

$$i\hbar \frac{\partial \psi_p(x,y,t)}{\partial t} = \left[-\frac{\hbar^2 \nabla^2}{2m_x} - \frac{\hbar^2 \nabla^2}{2m_y} + V(x,y)\right]\psi_p(x,y,t) + \frac{g_0}{2}\psi_e(x,y,t), \qquad (4)$$

$$i\hbar \frac{\partial \psi_e(x,y,t)}{\partial t} = E_{ex}\psi_e(x,y,t) + \frac{g_0}{2}\psi_p(x,y,t). \qquad (5)$$

Here, $\psi_p$ and $\psi_e$ represent the mean-field wave functions of photons and excitons, respectively. The effective masses of the cavity photons are $m_x = 2.0 \times 10^{-5} m_e$ and $m_y = 1.7 \times 10^{-5} m_e$ along $x$ and $y$ directions, where $m_e$ is the electron mass. Due to the heavy exciton mass, excitons are treated as dispersion-less, having a constant energy of $E_{ex} = 2.407$ eV, and the Rabi splitting energy $g_0$ is 100 meV. The geometry of the photonic potential $V(x, y)$ follows the polaritonic Dirac vortex lattice design, as illustrated in Fig. 1b. The potential values are 2213 meV inside the pillars and 2353 meV outside. The Hamiltonian corresponding to the CSEs is then diagonalized to determine the eigenstates of the system, as shown in Fig. 1d and 1e. To derive the band diagrams in Fig. 1c, we additionally apply the Bloch's theorem and the plane wave expansion method into this model.

**Acknowledgements**

R.S. and T. C. H. Liew gratefully acknowledge funding support from the Singapore Ministry of Education via the AcRF Tier 2 grant (MOE-T2EP50222-0008), AcRF Tier 3 grant (MOE-MOET32023-0003) "Quantum Geometric Advantage". R.S. also gratefully acknowledges funding supports from the Singapore Ministry of Education via Tier 1 grant (RG90/25) and Nanyang Technological University via a Nanyang Assistant Professorship start-up grant. R.S. and B. Z. acknowledge the Singapore National Research Foundation via a Competitive Research Program (grant no. NRF-CRP23-2019-0007).

**Author contributions**

R.S. and F. J. conceived the idea and design the research. F.J. and Y.X.L. synthesized the perovskite materials. F.J. fabricated the devices with the help of Z.Z. and J.H.R.. F.J. performed all the spectroscopy measurements. F.J. conceived the lattice model and H.Z. performed the theoretical calculations with inputs from T. C.H. L and B. Z.. R.S. and F.J. analyzed the data with inputs from T. C.H. L, D. S. and Q. Z.. R.S. and F.J. wrote the manuscript with input from all authors. R.S. supervised the research.

**Competing interests**

The authors declare that they don't have competing interests.

**Data availability**

All experimental data and code that support the plots within this paper are available from the corresponding author upon request.

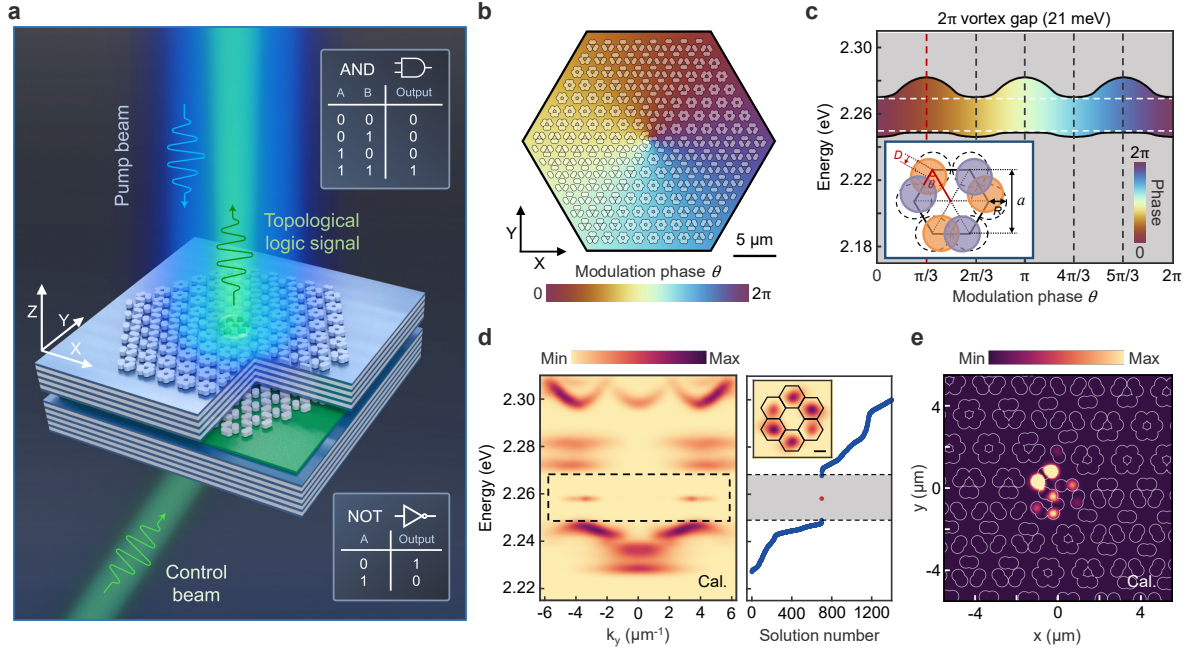

**Fig. 1: Scheme and design of the topological perovskite Dirac vortex microcavity. a**, Schematic illustration of dynamic topological exciton-polariton condensation with its logic functions, which are realized in a perovskite Dirac vortex microcavity. The pump beam and control beam are incident on the front and back side of the perovskite Dirac vortex microcavity, respectively, and the resulting topological logic signal can be detected from its front side. The inset of (**a**) is the operation criteria of the topological AND and NOT gates. **b**, The design of the polaritonic Dirac vortex lattice, and the color of the lattice illustrates the distribution of Kekulé modulation phase $\theta(r)$. **c**, The bandgap opening of different modulation phases $\theta$ (from 0 to $2\pi$) with the modulation amplitude of $D = 0.15a$, and the color represents the modulation phase. Inset of (**c**) is the detailed Kekulé modulation ($D = 0.15a$; $\theta = \pi/3$) in a hexagon supercell. **d**, Calculated polariton dispersion (left) and eigenstates (right) of the perovskite Dirac vortex lattice, revealing the topological Majorana-like state located exactly at mid-gap and spectrally isolated from all other trivial states. The inset of (**d**) is the calculated 2D momentum-space profile of the polaritonic topological Majorana-like state. Scale bar, 2 μm$^{-1}$. **e**, Calculated real-space profile of the polaritonic topological Majorana-like state demonstrating strong localization in single type sublattices at the vortex core.

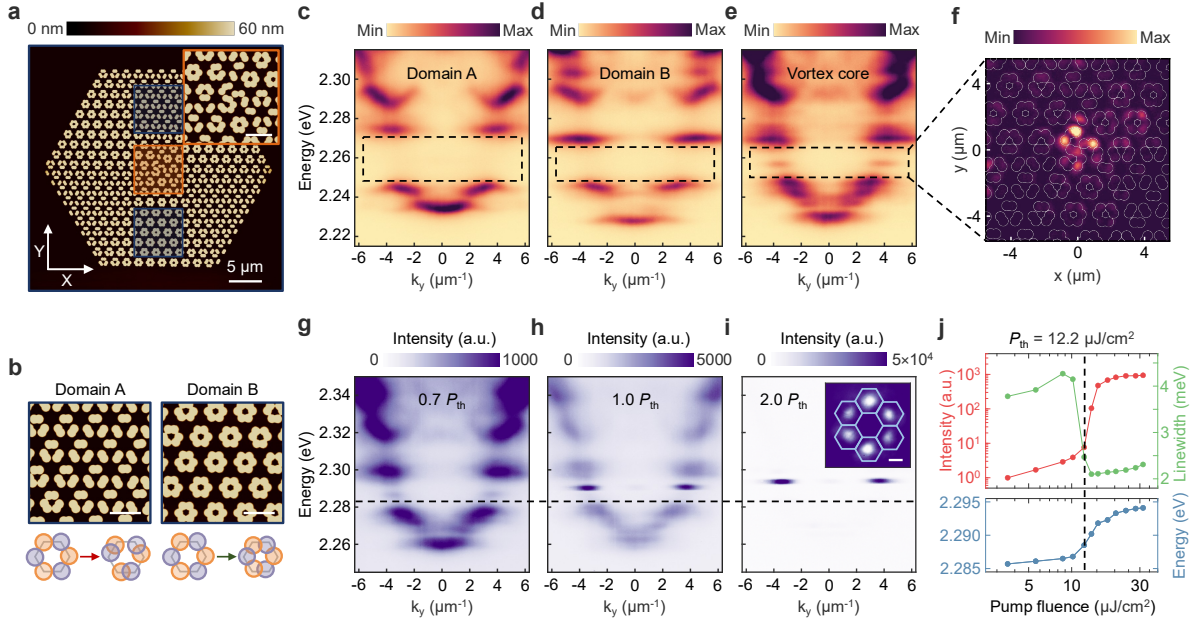

**Fig. 2: Experimental demonstrations of the polaritonic Majorana-like state with its spontaneous condensation. a**, Atomic force microscopy (AFM) image of the fabricated perovskite Dirac vortex lattice before the top DBR deposition, revealing the Kekulé modulation across the entire lattice profile. The blue and orange translucent overlays represent the bulk domains (near $\theta = \pi/3$ and $\theta = 4\pi/3$) and vortex core area, respectively. Inset of (**a**) is the magnified AFM image of the lattice core with a thickness of ∼ 60 nm. **b**, The magnified AFM images of the bulk domain A (near $\theta = \pi/3$, left) and bulk domain B (near $\theta = 4\pi/3$, right). The schematics of supercells illustrate the Kekulé modulation on the pristine honeycomb lattice. **c** and **d**, Experimental momentum-space polariton energy dispersions of the bulk domain A and domain B along the $k_y$ direction at $k_x = 0$ μm$^{-1}$, respectively. The black dashed lines highlight the bandgap opening. **e**, Experimental momentum-space polariton energy dispersion of the lattice core along the $k_y$ direction at $k_x = 0$ μm$^{-1}$. The black dashed lines highlight the emergence of the topological Majorana-like state in the middle of the topological gap. **f**, Experimental real-space photoluminescence image of the perovskite Dirac vortex lattice at energy of ∼ 2.257 eV, corresponding to the emission form the topological Majorana-like state. **g-i**, Momentum-space power-dependent dispersions of the polariton Dirac vortex lattice along the $k_y$ direction at $k_x = 0$ μm$^{-1}$ at $P = 0.7 P_{th}$ (**g**), $P = 1.0 P_{th}$ (**h**), $P = 2.0 P_{th}$ (**i**), respectively, which are non-resonantly pumped by a pulsed laser at 3.1 eV. Inset of (**i**) is the 2D momentum-space image of polariton condensation into the topological Majorana-like state at $P = 2.0 P_{th}$, where polaritons condense close to the $\Gamma$ points of the second BZs. Scale bar, 2 μm$^{-1}$. **j**, The evolution of integrated intensity and linewidth extracted from the topological state as a function of pump fluence, demonstrating a nonlinear increase in the integrated intensity and rapid decrease in linewidth at the threshold of $P_{th} = 12.2$ μJ/cm$^2$. Bottom panel of (**j**) is the peak energy evolution of the topological state during the condensation process, revealing a continuous blueshift.

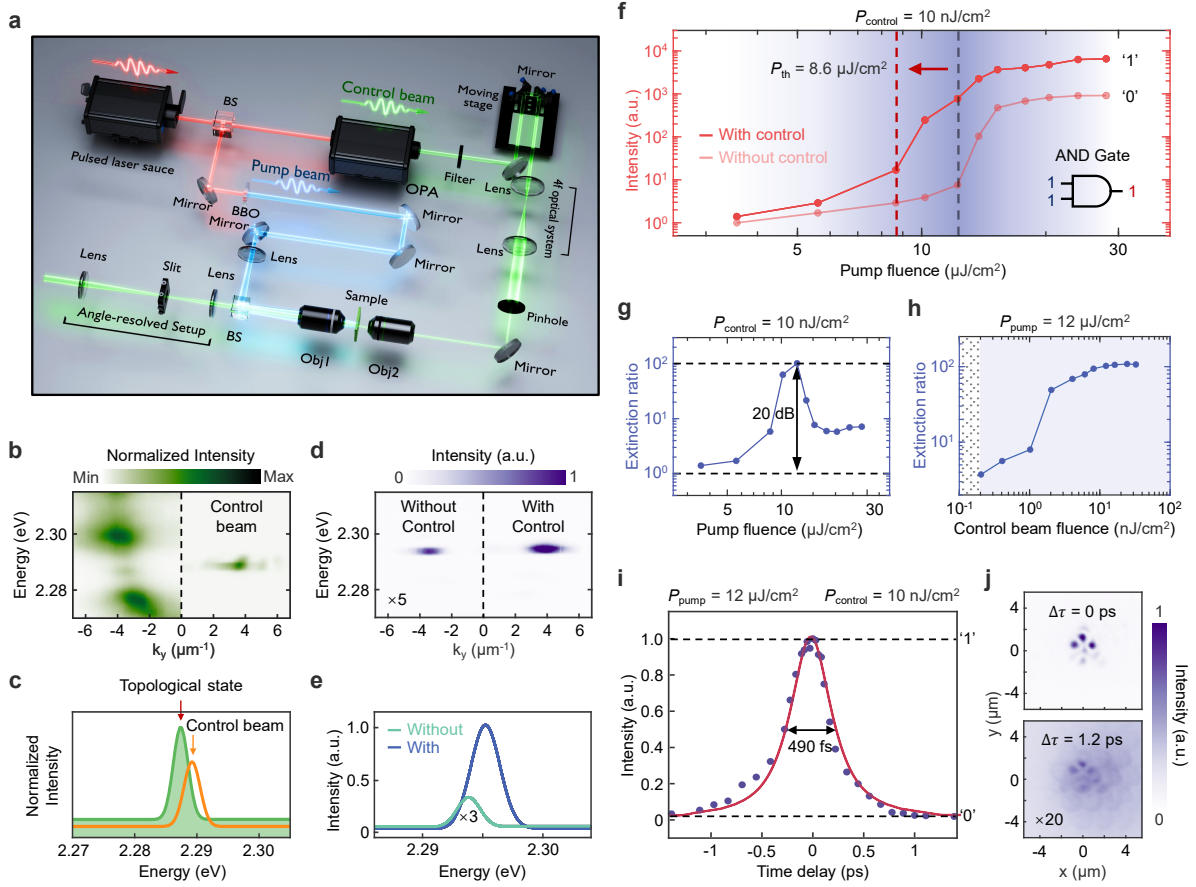

**Fig. 3: Dynamic amplification of topological exciton-polariton condensation enables topological AND gate.**
**a**, The schematic of the home-built optical measurement setup. A pulsed laser at 3.1 eV is normally incident on the front side of the sample, serving as pump beam. Another resonant pulsed laser at the target energy is injected at an oblique angle from the back of the sample, serving as control beam. **b**, Magnified polariton dispersion collected from the perovskite Dirac vortex lattice core at $k_x = 0$ μm$^{-1}$ (left), and the spectral position of the control beam resonant with the topological state (right). **c**, The integrated spectra of the topological state (green) and the control beam (orange) fitted by Gaussian functions, revealing the control beam nearly coincides with the topological state. **d**, Momentum-space dispersions of the topological exciton-polariton condensation without (left) and with (right) the synchronized control beam of 10 nJ/cm² under a pump fluence of 15 μJ/cm². The intensity under pump-only condition is multiplied by a factor of 5 for better comparison. **e**, The integrated spectra of the topological exciton-polariton condensation without (green) and with (blue) the synchronized control beam, fitted by Gaussian functions. **f**, The evolution of integrated intensity extracted from the topological state as a function of pump fluence for the pump-only condition (pink) and with the additional control beam seeding (red), corresponding to the logic level '0' and '1', respectively. **g**, The extinction ratio of the topological AND gate as a function of pump fluence under a fixed control fluence of 10 nJ/cm², revealing the highest extinction ratio of ~20 dB as the pump approaches the spontaneous condensation threshold. **h**, The dependence of extinction ratio on control beam fluence under a constant pump fluence of 12 μJ/cm². **i**, The integrated condensation intensities at different delay times, revealing an ultrafast response time of ~ 490 fs. **j**, Real-space photoluminescence images of the dynamic topological condensation at $\Delta\tau = 0$ ps (top) and $\Delta\tau = 1.2$ ps (bottom).

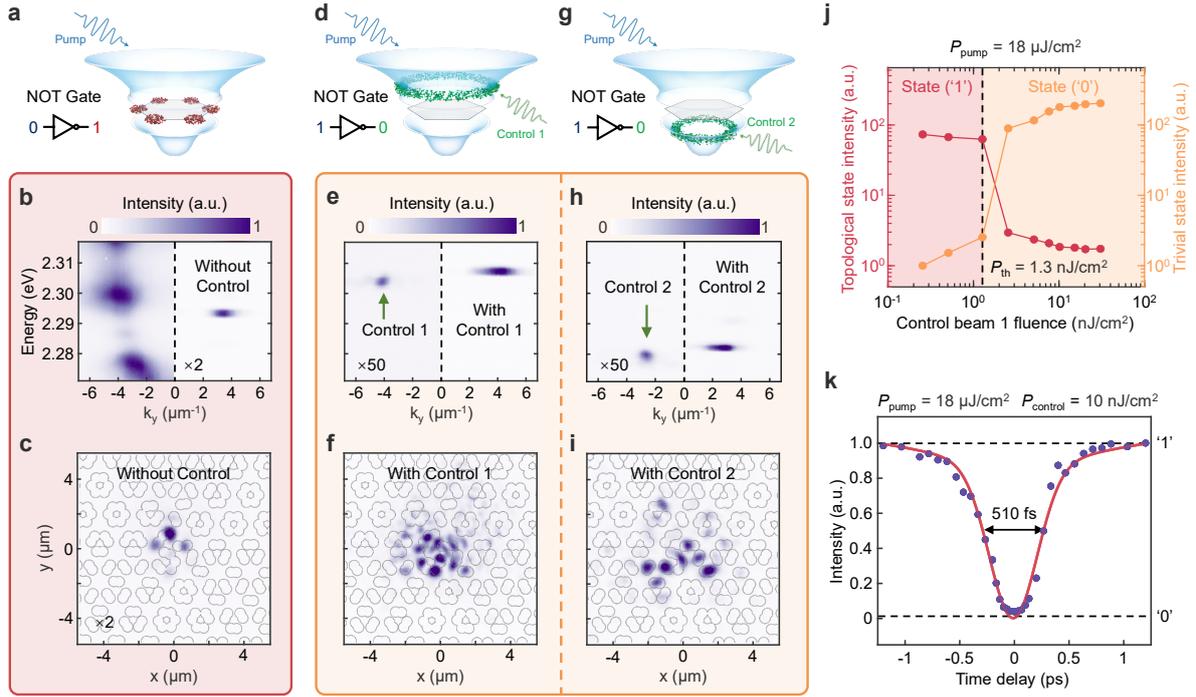

**Fig. 4: Towards topological exciton-polariton NOT gates. a**, The schematic of the spontaneous exciton-polariton condensation into the topological Majorana-like state in the absence of the control beam, which serves as the logic level '1' of the topological NOT gate. **b**, Magnified polariton dispersions collected from the Dirac vortex lattice core area at $k_x = 0$ μm$^{-1}$ below (left) and above (right) the condensation threshold, revealing the exciton-polaritons spontaneously condense into the topological Majorana-like state. **c**, The real-space photoluminescence profile of the topological Majorana-like state condensation. The intensity is multiplied by a factor of 2 for better comparison. **d** and **g**, The schematics of two distinct topological NOT gate configurations. When the control beams arrive, the spontaneous topological exciton-polariton condensation is quenched and diverted into the higher- (**d**) or lower-(**g**) energy trivial state condensation, corresponding to the logic level '0' of the topological NOT gate. **e** (left) and **h** (left), The spectral positions of the control beam 1 and control beam 2 resonant with a higher-energy and a lower-energy trivial states, respectively. Both control beams have the fluence of 10 nJ/cm², and they are multiplied by a factor of 50 for visibility. **e** (right) and **h** (right), The dispersions of the polariton condensation driven by the control beam 1 and the control beam 2 under the pump fluence of 18 μJ/cm², respectively, revealing that the topological condensation (logic level '1') are switched off, while the trivial states are massively occupied. **f** and **i**, The real-space photoluminescence profiles of the higher-energy (**f**) and lower-energy (**i**) trivial states polariton condensations driven by the control beam 1 and the control beam 2, respectively. **j**, The evolutions of the integrated intensity extracted from the topological state (red) and the higher-energy trivial state (orange) as a function of control beam 1 fluence at a pump fluence of 18 μJ/cm². **k**, The integrated topological condensation intensity of the topological NOT gate at different time delays between the pump beam and the control beam 1, and the FWHM of its temporal response is ~ 510 fs.

16